\documentclass[manuscript, sigconf, nonacm]{acmart}

\AtBeginDocument{%
  \providecommand\BibTeX{{%
    \normalfont B\kern-0.5em{\scshape i\kern-0.25em b}\kern-0.8em\TeX}}}

\usepackage{libertine}
\usepackage{natbib}
\usepackage{multirow}
\usepackage{booktabs}
\usepackage{amsmath}
\usepackage{graphics}
\usepackage{adjustbox}
\usepackage{url}
\usepackage{subfigure}
\usepackage{hyperref}
\usepackage{graphicx}
\usepackage{float}
\usepackage{csquotes}
\usepackage{xspace}
\usepackage{xcolor}
\usepackage{siunitx}
\usepackage{etoolbox}
  \robustify\bfseries

\usepackage{pifont}
\newcommand{\cmark}{\ding{51}}
\newcommand{\xmark}{\ding{55}}

\usepackage{caption}
\graphicspath{ {./Figures/} }

\DeclareUnicodeCharacter{03B3}{$\gamma$}
\newcommand{\CHMM}{C\lowercase{omorbidity}-HMM\xspace}
\newcommand{\A}{\textbf{\textsf{A}}\xspace}
\newcommand{\B}{\textbf{\textsf{B}}\xspace}
\newcommand\ie{i.\,e.\xspace}
\newcommand\eg{e.\,g.\xspace}

\definecolor{acute_red}{RGB}{222, 73, 104}
\definecolor{stable_blue}{RGB}{0, 91, 150}
\newcommand{\ACUTE}{\textcolor{acute_red}{acute}\xspace}
\newcommand{\STABLE}{\textcolor{stable_blue}{stable}\xspace}

\newcommand\blfootnote[1]{%
  \begingroup
  \renewcommand\thefootnote{}\footnote{#1}%
  \addtocounter{footnote}{-1}%
  \endgroup
}

\begin{document}

\title{Modeling Longitudinal Dynamics of Comorbidities}

\author{Basil Maag}
\affiliation{%
	\institution{ETH Zurich}
	\streetaddress{Weinbergstr. 56/58}
	\city{Zurich, Switzerland}
}
\email{basilmaag@gmail.com}

\author{Stefan Feuerriegel}
\affiliation{%
	\institution{ETH Zurich}
	\city{Zurich, Switzerland}
}
\email{sfeuerriegel@ethz.ch}

\author{Mathias Kraus}
\affiliation{%
	\institution{ETH Zurich}
	\city{Zurich, Switzerland}
}
\email{mathiaskraus@ethz.ch}

\author{Maytal Saar-Tsechansky}
\affiliation{%
	\institution{University of Texas at Austin}
	\city{Austin, Texas, USA}
}
\email{maytal.saar-tsechansky@mccombs.utexas.edu}

\author{Thomas Züger}
\affiliation{%
    \institution{Inselspital, Bern, University Hospital, University of Bern, Bern, Switzerland}
	\institution{ETH Zurich, Zurich, Switzerland}
}
\email{thomas.zueger@insel.ch}

\renewcommand{\shortauthors}{Basil Maag et al.}

\begin{abstract}
In medicine, comorbidities refer to the presence of multiple, co-occurring diseases. Due to their co-occurring nature, the course of one comorbidity is often highly dependent on the course of the other disease and, hence, treatments can have significant spill-over effects. Despite the prevalence of comorbidities among patients, a comprehensive statistical framework for modeling the longitudinal dynamics of comorbidities is missing. In this paper, we propose a probabilistic model for analyzing comorbidity dynamics over time in patients. Specifically, we develop a coupled hidden Markov model with a personalized, non-homogeneous transition mechanism, named \CHMM. The specification of our \CHMM is informed by clinical research: (1)~It accounts for different disease states (\ie, acute, stable) in the disease progression by introducing latent states that are of clinical meaning. (2)~It models a coupling among the trajectories from comorbidities to capture co-evolution dynamics. (3)~It considers between-patient heterogeneity (\eg, risk factors, treatments) in the transition mechanism. Based on our model, we define a spill-over effect that measures the indirect effect of treatments on patient trajectories through coupling (\ie, through comorbidity co-evolution). We evaluated our proposed \CHMM based on 675 health trajectories where we investigate the joint progression of diabetes mellitus and chronic liver disease. Compared to alternative models without coupling, we find that our \CHMM achieves a superior fit. Further, we quantify the spill-over effect, that is, to what extent diabetes treatments are associated with a change in the chronic liver disease from an acute to a stable disease state. To this end, our model is of direct relevance for both treatment planning and clinical research in the context of comorbidities.
\end{abstract}




\maketitle

% DOI
\blfootnote{\url{https://doi.org/10.1145/3450439.3451871}}

\section{Introduction}

In medicine, comorbidities refer to the simultaneous occurrence of several diseases affecting patient health \citep{ComorbidityMultimorbidity}. Recent estimates suggest that 50\,\% of hospitalizations in the U.S. involve patients where eleven and more different disease codes are present \citep{vanweenen2020estimating}. Similarly, multiple diseases and, especially, chronic ones simultaneously affect the health of a large proportion of elderly patients \citep{BARNETT201237, WolffPrevalenceChronic}. This calls for a better understanding regarding the interaction of diseases, that is, the dynamics behind comorbidities.

Comorbidities are characterized by a concomitant or concurrent evolution, whereby the course of one disease is highly dependent on the course of the other disease \citep{ComorbidityMultimorbidity}. An example are chronic pain and depression. Both often co-occur and their evolution is further highly dependent. Chronic pain often causes a depression and, vice versa, a depression increases typically the perceived level of pain \citep{BairMatthewJ2003DaPC, Geerlings:2002aa}. Another common example of comorbidities are diabetes and chronic liver diseases (\eg, non-alcoholic fatty liver disease) \citep{Tolman734, Harris115, HICKMAN2007829}. The evolution of both diseases is known to be subject to profound interactions and metabolic alterations. This is also reflected in a recent proposal to change the nomenclature from non-alcoholic fatty liver disease to metabolic associated fatty liver disease (MAFLD) \citep{ESLAM20201999}. To this end, a better understanding of the co-evolution could inform treatment planning and is thus of direct clinical relevance. However, this requires models where the co-evolution is considered, that is, where disease interactions over time are subject to statistical modeling. 
		
Statistical models that are tailored for comorbidities are scarce. One stream of literature has studied which diseases are often found together \citep{Guisado-Clavero:2018aa, huang2017longitudinal, Violan:2020aa}. However, this question relates to unsupervised learning (\eg, clustering) and does not analyze the longitudinal interactions of the diseases. Another stream of literature is concerned with the impact of comorbidities on patient health \citep[\eg,][]{CarnethonLongitudinal, Geerlings:2002aa, RICHARDSON2008509, Terzano:2010aa,mueller2020longitudinal} or readmission risk \citep{vanweenen2020estimating}. However, in these works, comorbidities are merely used as independent variables but not as dependent variables, that is, outcomes of interest. Because of this, such models cannot describe the co-evolution among comorbidities. This motivates our research: \emph{how does treating disease \A affect the course (disease state) of disease \B?} To answer this question, statistical inferences are needed that allow for \textbf{interpretability}.

In medicine, the course of diseases is commonly classified by different trajectory phases called \textquote{disease states}, whereby diseases undergo different phases of \emph{\ACUTE} and \emph{\STABLE} states. These disease states underlie decision-making in a clinical practice as treatment plans and regimens differ depending on whether a disease is in an \ACUTE or in a \STABLE state (cf. the Corbin-Strauss trajectory framework \citep{Strauss:1988aa}). However, the disease states cannot be directly observed \citep{Corbin:1991aa}. Specifically, lab measurements provide only noisy realizations of the disease state and, therefore, both measurements (\eg, glucose levels or even other forms of symptoms) and disease states are related only stochastically. To reflect this, a common approach is to model health trajectories in a way that disease states are formalized through the use of latent states and thereby represent hidden Markov models~(HMMs) \citep[\eg,][]{SukkarDiseaseprogression, ShirleyALCOHOLISM, oezyurt2021attdmm, MartinoHMMdisease, BartolomeoNicola2011Polc, ScottHMMLongitudinal}. In this work, we adhere to the so-called trajectory framework \citep{Strauss:1988aa,Corbin:1991aa} and describe the evolution of disease based on disease states, yet we extent the HMM-based framework from a single-disease to a multi-disease setting.

\textbf{Model:} In this work, we develop a probabilistic model for analyzing the longitudinal progression of comorbidities in patients. Specifically, we propose a tailored coupled hidden Markov model (CHMM) with a personalized, non-homogeneous transition mechanism, named \CHMM. Informed by clinical practice, we specify the \CHMM as follows: (1)~We account for different states (\ie, \ACUTE, \STABLE) in the disease progression by introducing latent states that are of clinical meaning. (2)~We model a coupling among the trajectories from comorbidities to capture co-evolution dynamics. (3)~We consider between-patient heterogeneity (\eg, risk factors, treatments) in the transition mechanism. We further provide a fully Bayesian estimation framework through Markov chain Monte Carlo (MCMC). Based on our model, we formalize and estimate a spill-over effect which quantifies the indirect effect of treatments through the co-evolution of the diseases. This allows for statistical inferences that ensure interpretability. 

\textbf{Findings:} Our \CHMM is evaluated based on a large longitudinal dataset. Specifically, we leverage electronic health records from 675 patients over the course of 10 years. The electronic health records document the joint progression of (\A)~diabetes and (\B)~chronic liver disease. Patients at risk of developing diabetes mellitus type 2 were chosen as this disease is among the most common diseases worldwide and its presence is often associated with a chronic liver disease as a concomitant comorbidity \citep{Tolman734, Harris115, HICKMAN2007829}. We obtain three main findings. First, modeling disease interactions through means coupling results in a superior model fit. Second, an \ACUTE state in one of the two diseases is associated with an increased probability of transitioning to an \ACUTE state in the other disease. Third, a treatment targeting diabetes has considerable spill-over effects: it is not only associated with a decrease of the risk of an \ACUTE disease state in diabetes but also in chronic liver disease. Altogether, the results demonstrate empirically the importance of modeling the co-evolution among comorbidities.

\textbf{Contributions:}\footnote{Our code is available via \url{https://github.com/mb2019/Comorbidity-HMM}} Our work advances existing research on modeling longitudinal disease dynamics in the following ways.
\begin{enumerate}
\item To the best of our knowledge, our \CHMM is the first statistical model that is specifically tailored to capture longitudinal dynamics among comorbidities. 
\item For this purpose, we develop a coupled hidden Markov model with a non-homogeneous transition mechanism for longitudinal data and, by deriving the likelihood, provide a Bayesian estimation procedure. 
\item We formalize and estimate a spill-over effect that measures the indirect effect of treatments through disease interactions. This allows for direct interpretability as it quantifies empirically how treating disease~\A affects the course (disease state) of disease~\B. Thereby, we generate new knowledge for clinical practice (\eg, treatment planning) and research.
\end{enumerate}

\section{Related Work}
\label{sec:related_work}

\subsection{Comorbidities}

In medicine, the term \textquote{comorbidity} refers to co-occuring diseases (see \citep{ComorbidityMultimorbidity}). In the literature, comorbidities are sometimes classified into a primary disease and secondary diseases, yet such a classification is not universally applicable \citep{FEINSTEIN1970455, Valderas357}. Hence, we later make no such distinction and simply refer to disease~\A and disease~\B. Both diseases may simply co-occur by chance, yet they might also be subject to a causal relationship or due to correlated or shared risk factors \citep{Valderas357}. This means that interactions among diseases occur. Specifically, by treating one disease, one oftentimes can also expect a better course of co-occurring diseases. 

Comorbidities are widespread in clinical practice. In particular, patients who are hospitalized often suffer from multiple comorbidities (\eg, every second hospitalization lists eleven or more different disease codes \citep{vanweenen2020estimating}). Due to their prevalence, comorbidities are also of direct relevance for clinical decision-making. In particular, the presence of comorbidities might worsen health outcomes and, hence, clinical practice emphasizes that their treatment should be given special consideration \citep{GIJSEN2001661}. 

In clinical practice, there are numerous examples of comorbidities. For instance, depression and chronic pain occur often jointly, especially their severity is often dependent \citep{BairMatthewJ2003DaPC, Geerlings:2002aa}. Similarly, up to 40\,\% of cancer patients also suffer from psychological distress such as anxiety or depression \citep{Zabora:2001aa}. Further, patients with diabetes are also frequently diagnosed with additional comorbidities \citep{IglayComorbidities}, such as hypertension, cardiovascular, or, as studied in this paper, chronic liver diseases (\eg, non-alcoholic fatty liver disease) \citep{Tolman734, Harris115, HICKMAN2007829}.

\subsection{Statistical Inferences using Comorbidity Data}

Despite the importance of comorbidites in medicine, there are only few studies that have incorporated their structure into the model. We provide a summary of key references below. While they also use data on comorbidities, their objective is different from our work.

One stream in the literature has used statistical models to detect co-occurrence patterns among diseases \citep[\eg,][]{Guisado-Clavero:2018aa, huang2017longitudinal, Violan:2020aa}. This allows to make associations of which diseases co-occur and should thus be regarded as comorbidities. However, these studies address the question of which comorbidities appear frequently together but they do not model their course. For this different models can be used. For instance, comorbidities can be represented as networks in which nodes correspond to symptoms and edges to potentially causal relationships \citep{cramer_waldorp_2010}. Based on the network, shared or overlapping symptoms of distinct disorders then indicate the onset of comorbidities. This hypothesis has also been tested using dynamic structural equation models \citep{Groen:2020aa, BringmannLongitudinalNetwork} or deep diffusion processes \citep{qian2020learning}. This answers the question of \emph{which} diseases co-occur but not how, that is, without providing insights into longitudinal disease interactions. Similarly, Bayesian networks have been used to investigate comorbidities and their temporal dependencies \citep{FaruquiBayesianNetwork, LAPPENSCHAAR2013171, LAPPENSCHAAR20131405}. However, these works offer only little insights into the actual disease progression such as underlying disease states.  

Other works use data on comorbidities as input when making inferences regarding patient outcomes \citep[\eg,][]{CarnethonLongitudinal, Geerlings:2002aa, RICHARDSON2008509, Terzano:2010aa, ZOLBANIN2015150} including readmission risk \citep{vanweenen2020estimating}. This allows to control for the fact that the co-occurrence of diseases often introduces associations with patient outcomes that are of nonlinear form. For instance, patient survival might be lowered when there are many co-occurring diseases. In clinical practice, this is reflected in different measures such as, \eg, the Charlson comorbidity index \citep{Charlson:1987aa}. However, the aforementioned works consider comorbidities as independent variables and not as dependent variables.

\subsection{Hidden Markov Models}

Hidden Markov models (HMMs) are a flexible statistical technique for modeling time series. HMMs model a sequence of latent states that follow a Markov chain, based on which, observations are emitted \citep{GhahramaniZoubin2001AITH}. HMMs have found widespread application in modeling human behavior. Examples include clickstream data \citep{HMMClickstream}, urban activities \citep{HMMUrbanDynamic}, and hiring decisions \citep{KokkodisHiring}.

HMMs have previously been applied to data on patient health \citep[\eg,][]{MacKayAltmanMixedHMM, JacksonHMM, oezyurt2021attdmm, JacksonMultistateMarkov, SukkarDiseaseprogression, ScottHMMLongitudinal}. In these works, HMMs are further adapted in order to reflect patient trajectories through following specification. First, the latent states commonly correspond to different diseases states (\ie, phases) in the trajectory \citep[see][]{Corbin:1991aa}. Second, covariates as well as random effects are incorporated in the transition mechanism in order to account for patient heterogeneity and other sources of unobserved heterogeneity \citep{MacKayAltmanMixedHMM}. However, the aforementioned HMMs describe the evolution of single-disease outcomes but not the evolution in multi-disease settings, that is, comorbidities.

There exist several variants of HMMs \citep[\eg,][pp.\,76--77]{Sucar2015PGM}. For instance, coupled HMMs are an extension of HMMs that could potentially model the evolution of multiple interacting processes. These CHMMs allow the latent states of distinct but possibly correlated processes to be conditional dependent \citep{BrandCoupledHMM}. Here the latent states of different sequences can be linked using a Cartesian product, which gives the basis of a so-called Cartesian product CHMM. In a simulation study, such a Cartesian product CHMM was shown to outperform ordinary and multivariate HMMs when the underlying processes are dependent \citep{PohleJennifer2020Apoc}.

CHMMs have been applied in various areas, including speech recognition \citep{NefiancoupledHMM}, epileptic seizure detection \citep{CraleyEpileptic}, and sleep staging \citep{RezekEstimation}. Furthermore, they have been used to model the spreading of infectious diseases \citep{PanayiotaBayesian, dong2012graphcoupled}. In \citep{PohleJennifer2020Apoc}, a Cartesian product CHMM is developed for modeling the joint progression of lab measurements in a single-disease setting from intensive care units. In ecology, a CHMM was utilized to investigate the interdependence among voles from different forest regions \citep{SherlockChris2013AchM}. More specifically, it was examined if the occurrence of one disease makes the contagion of another one more likely. Further, a multi-chain Markov switching model akin to a CHMM was employed to check if there is volatility spillover between geographical financial markets \citep{GALLO20083011}.

\textbf{Research gap:} The importance of comorbidities for clinical practice has been widely recognized, and yet statistical frameworks for modeling their co-evolution over time have received little attention.
However, understanding the underlying disease dynamics could be of significant value for treatment planning. To this end, we develop a \CHMM that is tailored to capture comorbidity dynamics. Specifically, we model the disease interactions among comorbidites in patients based on longitudinal data.

\section{Proposed Model: \CHMM}
\label{modeldevolpment}

\subsection{Problem Statement}

We aim to model the co-evolution of comorbidities. For this, we make use of a longitudinal dataset with health information from patients. We further consider two diseases as given by disease~\A and disease~\B. Here we explicitly assume that the evolution of both is characterized by disease interactions. This motivates a better understanding of \emph{how} the course of both diseases interact. Specifically, we seek to answer the question whether there are potential spill-over effects: \emph{how does treating disease~\A affect the state of disease~\B?} For instance, to what extent does Metformin as diabetes treatment influence the risk of an \ACUTE state of a co-occurring liver disease? Formally, we want to measure the influence of one disease on the state of the other. We thus develop a model that ensures interpretability in order to allow for such statistical inferences.

To be of clinical relevance, our model specification accommodates further properties, namely (1)~disease states, (2)~coupling, and (3)~between-patient heterogeneity, as described in the following.

\vspace{0.2cm}
\underline{(1)~Disease states}. The course of many diseases (especially chronic or otherwise long-lasting diseases) undergoes different phases over time, where the disease is in an \textquote{\ACUTE} or \textquote{\STABLE} phase \citep{Corbin:1991aa}. In medicine, the phases are typically termed \textquote{disease states}. These states themselves cannot be directly identified and, instead, can only be recovered from measurements \citep{Corbin:1991aa}. To accommodate this, we follow prior literature \citep[\eg,][]{SukkarDiseaseprogression, ShirleyALCOHOLISM, oezyurt2021attdmm, MartinoHMMdisease, BartolomeoNicola2011Polc, ScottHMMLongitudinal} and model disease states through latent states. We later name the latent states \textquote{\ACUTE} and \textquote{\STABLE} to reflect their clinical meaning.  Consistent with prior literature \citep[\eg,][]{SukkarDiseaseprogression, ShirleyALCOHOLISM, oezyurt2021attdmm, MartinoHMMdisease, BartolomeoNicola2011Polc, ScottHMMLongitudinal}, we thus build upon the HMM-based framework \citep{RabinerTutorial} for modeling the course of diseases \A and \B. Intuitively, we represent both through two separate HMMs. 

\vspace{0.2cm}
\underline{(2)~Coupling}. We expect disease interactions among comorbidities and, therefore, explicitly assume the states of disease \A and \B to be dependent. As an example, when a disease~\A transitions from a \STABLE to an \ACUTE state, it should also increase the probability of disease~\B to move from a \STABLE to an \ACUTE state. For this reason, we later introduce a coupling between the HMMs from disease \A and \B. 

\vspace{0.2cm}
\underline{(3)~Between-patient heterogeneity}. Disease dynamics are known to vary extensively across patients \citep[\eg,][]{MuslinerHeterogeneity, Walraven:2015aa}. In medicine, this is described via risk factors (\eg, age, gender) and the presence of treatments \citep[\eg,][]{hatt2021estimating}. Therefore, our model accommodates between-patient heterogeneity as follows. On the one hand, we incorporate patient-level covariates that denote risk factors and treatments. These are entered in the transition mechanism of the HMMs, so that risk factors (as well as treatments) affect the disease dynamics: this allows risk factors (as well as treatments) to increase the propensity of \ACUTE vs. \STABLE disease states.\footnote{The choice that risk factors (and treatments) should enter not the emission but the transition mechanism is due to clinical research. The reasoning is the following: variables in the emission only affect observed measurements (\eg, they affect pain resistance) as they moderate how \ACUTE vs. \STABLE disease states link to measurements; yet they cannot affect the course of the disease progression. This choice is later validated in our numerical experiments.} On the other hand, we account for unobserved heterogeneity. Here we follow prior literature \citep{bell_jones_2015, CurranDisaggregation, EndersCraigK.2007CPVi, HamakerCenter, HoffmanPersonsasContexts,WangLijuanPeggy2015ODBa} and control for within effects in the treatment variable (see Sec.~\ref{sec:model_variables} for details).   

\vspace{0.2cm}

In the following, we develop a model named \CHMM. Our \CHMM accommodates the above properties (1)--(3). Based on the \CHMM, we define a spill-over effect which estimates the influence of a treatment from disease~\A on the state from disease \B. Here we follow clinical guidelines that base treatment planning primarily on disease states instead of measurements or symptoms \citep{Strauss:1988aa,Corbin:1991aa, larsen2017lubkin}. For that reason, we define the spill-over effect with regard to disease states as opposed to measurements or symptoms. This allows us to make statistical inferences of how a treatment targeting disease~\A render the risk of disease~\B to become \ACUTE vs. \STABLE. 

\subsection{Model Specification}
Our \CHMM consists of the following five components: (i)~latent states representing the disease states; (ii)~observations representing the measurements or symptoms; (iii)~an emission component linking states and observations; (iv)~a transition mechanism; and (v)~a coupling. Based on it, we model the evolution for each of two diseases \A and \B through a HMM-based framework that is additionally subject to coupling. The components (i)--(v) are specified in the following:

\vspace{0.3cm}
\noindent\textbf{(i)~Observations:} 
The observations denote measurements (\eg, from labs or data on symptoms) for the two diseases \A and \B. Let $i= 1, \ldots, N$ denote the different patients. Further, let $t = 1, \ldots, T$ denote the different time steps. Then observations of the diseases \A and \B are given by $Y_{it}^{(A)}$ and $Y_{it}^{(B)}$, respectively. 

\vspace{0.3cm}
\noindent\textbf{(ii)~Latent states:} The latent states represent the different disease states (\ie, \ACUTE vs. \STABLE). Here we introduce latent states $S_{it}^{(A)} \in \mathcal{S}^{(A)}$ and $S_{it}^{(B)} \in \mathcal{S}^{(B)}$ for patient $i$ at time step $t$, where the latent states belong to disease \A and \B, respectively. As such, each disease has a separate latent state and can thus attain a different state. For each disease, we assume two different states \textquote{1} and \textquote{2}, \ie, $\mathcal{S}^{(A)} = \mathcal{S}^{(B)} = \{ 1, 2 \}$. For ease of notation, we refer to state $1$ as \textquote{\STABLE} and state $2$ as \textquote{\ACUTE}. In total, this allows for $\big\vert \mathcal{S}^{(A)} \big\vert \times \big\vert \mathcal{S}^{(B)} \big\vert$ different combinations.

\begin{table}[H]
\centering
\caption{Mapping of latent states.} \label{tbl:state_mapping}
\begin{tabular}{cc cc}
\toprule
\multicolumn{2}{c}{Disease states} &  Latent states & Global state \\
\cmidrule(lr){1-2}
 Disease \A & Disease \B & $\big(S_{it}^{(A)}, S_{it}^{(B)} \big)$ & ${S_{it}^{(G)}}$ \\
\midrule
\STABLE & \STABLE & (1,1) & 1 \\
\STABLE & \ACUTE  & (1,2) & 2 \\
\ACUTE & \STABLE  & (2,1) & 3 \\
\ACUTE & \ACUTE   & (2,2) & 4 \\
\bottomrule
\end{tabular}
\end{table}

We now define a global latent state based on the Cartesian product of the two separate latent states. Formally, we define $S_{it}^{(G)}=\big(S_{it}^{(A)}, S_{it}^{(B)}\big)$. This is later necessary when we detail the coupling as part of component~(v). The underlying mapping is listed in Tbl.~\ref{tbl:state_mapping}. For instance, a global state $S_{it}^{(G)} = (2,2)$ means that both diseases are in an \ACUTE state. As a short form, we use the notation $S_{it}^{(G)} \in \mathcal{S}^{(G)} = \{1, \ldots, 4\}$.

\vspace{0.3cm}
\noindent\textbf{(iii)~Emission component:} The emission component defines the probability of an observation conditional on the latent state. In our case, it defines the probability of observing a measurement or symptom conditional on disease state. For this, both observations $Y_{it}^{(A)}$ and $Y_{it}^{(B)}$ of the two diseases are assumed to depend on the latent states $S_{it}^{(A)}$ and $S_{it}^{(B)}$, respectively. Given a latent state $s$, we model them to be conditionally normally distributed via
\begin{equation}
\begin{split}
Y_{it}^{(A)} \;\bigm|\; S_{it}^{(A)} = s \sim \mathcal{N}\Big(\mu_s^{(A)}, \sigma^{2}_{(A)} \Big), \\
Y_{it}^{(B)} \;\bigm|\; S_{it}^{(B)} = s \sim \mathcal{N}\Big(\mu_s^{(B)}, \sigma^{2}_{(B)} \Big) \vphantom{,}
\end{split}
\end{equation}
with mean $\mu_s^{(A)}$, $\mu_s^{(B)}$ and variance $\sigma^2_{(A)}$, $\sigma^2_{(B)}$. Here the observations are modeled with an individual mean per state $s$ and disease. Hence, an \ACUTE state from disease \A might have, on average, larger values as observed measurements or symptoms. The variance is disease-specific but independent of the state. In the above specification, we used a normal distribution to model the observations as, in our experiments, the measurements (\eg, blood glucose levels) follow a normal distribution. 

\vspace{0.3cm}
\noindent\textbf{(iv)~Transition mechanism:} The transition mechanism specifies the probability of moving from one latent disease state to another latent disease state. As typical for HMMs, we assume the latent states to follow a Markov process. However, we accommodate additional covariates (\ie, risk factors, treatments) whereby we yield a non-homogeneous transition mechanism that is personalized to patient profiles.

Formally, we model the probabilities of the global states via a multinomial logit link function \citep{MacKayAltmanMixedHMM, Leos-BarajasVianey2018AItA}. Hence, the transition probability $\gamma^{(j\rightarrow k)}_{it}$ denotes the probability of patient $i$ transitioning from global state $j$ at time $t$ to global state $k$ at time $t+1$. We specify the transition probability via
\begin{equation}
\begin{split}
\gamma^{(j\rightarrow k)}_{it} & = P\Big(S_{i,t+1}^{(G)}=k \;\bigm|\; S_{it}^{(G)} =j \Big) \\
& = \frac{\exp(\eta_{jk})}{\sum_{l=1}^{\vert \mathcal{S}^{(G)} \vert}\exp(\eta_{jl})}, 
\end{split}
\end{equation}
where
\begin{equation} 
\label{eqn:tpm_eta}
\eta_{jk}=
\begin{cases}
\alpha_{jk} + x_{it}^T\beta_{jk} , & \textrm{if } j \neq k, \\
0 , & \textrm{if } j=k.
\end{cases}
\end{equation}
The above logit makes use of further variables: The variable $a_{jk}$ in Eq.~\ref{eqn:tpm_eta} corresponds to the intercept of the transition from global state $j$ to $k$. All else equal, it captures how likely a certain disease state transition is. The covariates $x_{it}$ describe the between-patient heterogeneity (\eg, risk factors, treatments). The coefficients of the covariates $x_{it}$ are given by $\beta_{jk}$. These capture an elevated probability of \ACUTE states (over \STABLE states) due to risk factors. Covariates from the treatment are further processed to control for within effects in the treatment variable (see Sec.~\ref{sec:model_variables}). Coefficients for $j = k$ are set to zero in order to ensure identifiability. Hence, transitions where the same latent disease state is maintained (\ie, recurrent transitions) serve as the reference category; see Tbl.~\ref{tbl:tpm}.

The initial probability  of latent state $s$ is given by $\pi_s$ (\ie, the probability of latent states in time step $t = 1$). The initial probabilities were estimated independently of each patient and any risk factors. This approach follows previous works \citep{WangLongitudinalHMM}.

\begin{figure}[H]
\centering
\includegraphics[width=\columnwidth]{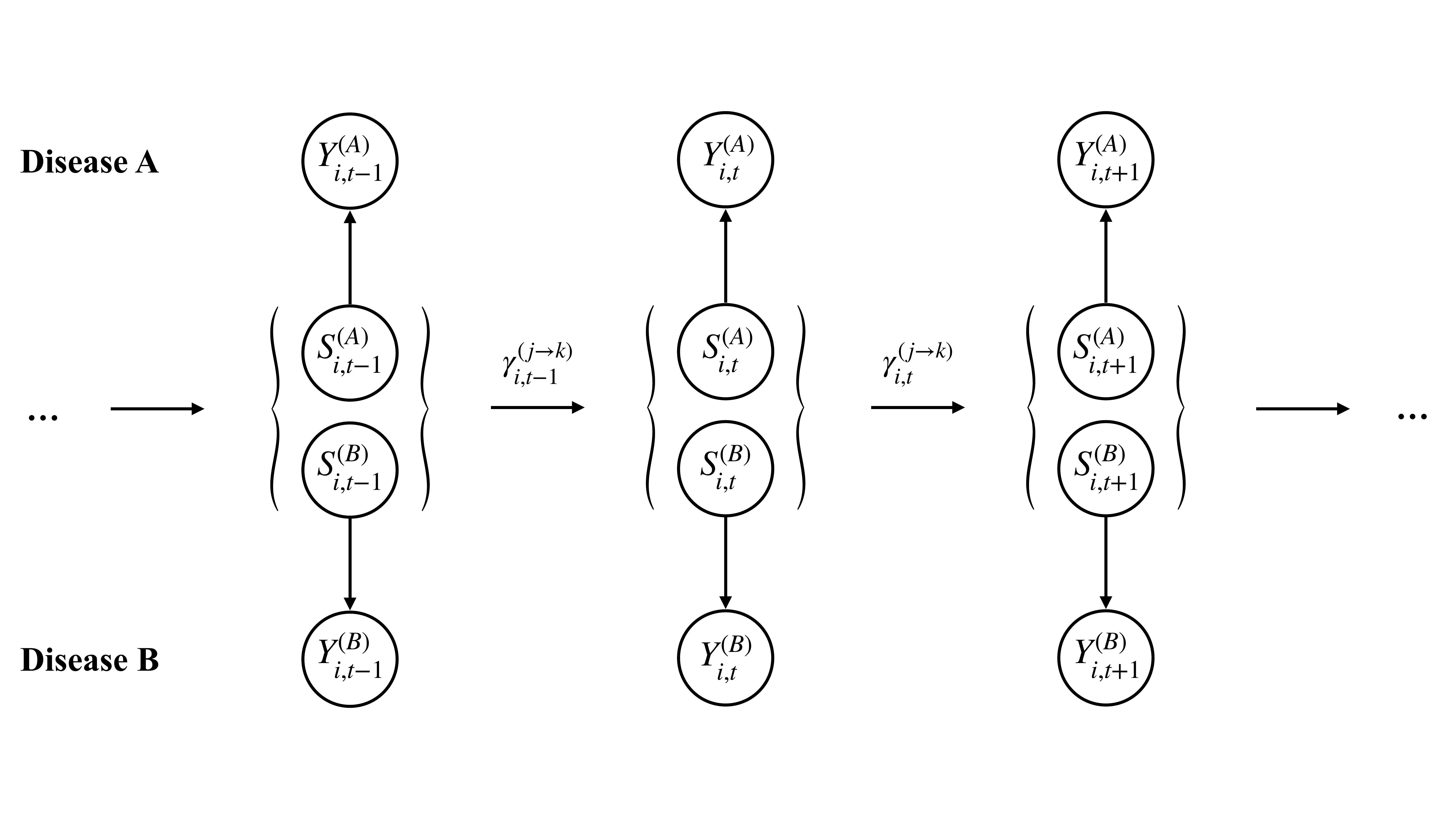}
\caption{Illustrative scheme of coupling inside the \CHMM.}
\label{fig:CPCHMM}
\end{figure}

\vspace{0.3cm}
\noindent\textbf{(v)~Coupling:} We model the interaction of the two diseases by introducing a coupling among their latent states. Specifically, we build upon the concept of a Cartesian product coupled HMM \citep{PohleJennifer2020Apoc}. For this, we draw upon the Cartesian product of the underlying latent states from above. Recall that $S_{it}^{(G)}=(S_{it}^{(A)}, S_{it}^{(B)})$, while $S_{it}^{(G)}$ is referred to as global state. Then we assume that the global states follow a Markov chain (Fig.~\ref{fig:CPCHMM}).

Mathematically, by introducing the global states, the coupling is captured as part of the transition mechanism. The reason is that the transition mechanism is now based on the global state, so that the probability of $S_{it}^{(A)}$ does not only depend on the previous latent state $S_{i,t-1}^{(A)}$ but also on that from the other disease $S_{i,t-1}^{(B)}$. The same holds true analogously when exchanging disease \A and \B. Hence, the transition mechanism accounts for potential co-movements from the states of both diseases. 

The coupling among the latent states is best seen in an example. Recall that the transition mechanism depends on an intercept but it is also influenced by further covariates (\ie, risk factors). We now discuss how the transitions of one disease differ depending on the other diseases. For instance, it might be the case that the probability of a transition from a \STABLE to an \ACUTE state for disease \B is more likely given that disease~\A is in an \ACUTE state. Such a case could, for instance, arise if the corresponding intercept $\alpha_{34}$ [for transition (2=\ACUTE,1=\STABLE) $\to$ (2=\ACUTE,2=\ACUTE)] is substantially larger than the intercept $\alpha_{12}$ [for transition (1=\STABLE,1=\STABLE) $\to$ (1=\STABLE,2=\ACUTE)]. This means that an \ACUTE state for disease \A is associated with an increased propensity of disease~\B also moving to an \ACUTE state (rather than a \STABLE state). Thus, comparing the estimated intercepts of such transitions can offer insights about the underlying disease interactions. Hence, we later report our estimations results for $\alpha_{jk}$ as this indicates the strength of the underlying coupling, \ie, the underlying disease interaction.

\subsection{Spill-Over Effect}

Treatments such as medications can also be included in the covariates. If such a treatment is specifically targeted at one comorbidity, it will have a direct effect on the state transitions of that disease. For instance, a drug designed for disease \A may prevent the transition from state $(1,1)$ to $(2,1)$ (\ie, prevent \A from becoming \ACUTE). However, the same drug might also implicitly effect the state transitions of disease \B through their coupling. We will therefore refer to the former as direct treatment effect and to the latter as spill-over effects. 

An example of the a spill-over effect is the following. Let us consider the state transitions $(2,2) \rightarrow (1,2) \rightarrow (1,1)$. It represents the scenario in which a treatment makes disease \A transition from an \ACUTE state to a \STABLE one (\ie, direct effect). Subsequently, disease \B also transitions to a \STABLE state in the following time period (\ie, as an additional indirect effect). 

We formalize the above spill-over effect as follows. Let $z$ denote a treatment which is targeted at disease \A and affects the transition probabilities via the covariates $x_{it}$ as denoted in Equation \ref{eqn:tpm_eta}. Then, the probability of the aforementioned state transitions of patient $i$ under treatment $z$ is given by 
\begin{equation}
\begin{split}
\xi_{it}(z) & = \gamma_{it}^{(4 \rightarrow 2)} \times \gamma_{i,t+1}^{(2 \rightarrow 1)} \\
& = P\Big(S_{i,t+1}^{(G)}=2 \;\bigm|\; S_{it}^{(G)} =4,  Z=z \Big) \\
& \times P\Big(S_{i,t+2}^{(G)}=1 \;\bigm|\; S_{i,t+1}^{(G)} =2,  Z=z \Big). \\
\end{split}
\end{equation}
Analogously, the probability under no treatment denoted with $z^\prime$ is given by
\begin{equation}
\begin{split}
\xi_{it}(z^\prime) & = {\gamma_{it}^{(4 \rightarrow 2)}}^\prime \times {\gamma_{i,t+1}^{(2 \rightarrow 1)}}^\prime \\
& = P\Big(S_{i,t+1}^{(G)}=2 \;\bigm|\; S_{it}^{(G)} =4,  Z=z^\prime \Big) \\
& \times P\Big(S_{i,t+2}^{(G)}=1 \;\bigm|\; S_{i,t+1}^{(G)} =2,  Z=z^\prime \Big). \\
\end{split}
\end{equation}
Thus, if the treatment $z$ increases the probability $\xi_{it}(z)$ but is specifically designed for disease \A, it also affects disease \B positively through their coupling. Hence, we refer to 
\begin{equation}
\text{SpillOver}_{(2,2) \rightarrow (1,2) \rightarrow (1,1)} := \xi_{it}(z) - \xi_{it}(z^\prime) 
\end{equation}
as spill-over effect for $(2,2) \rightarrow (1,2) \rightarrow (1,1)$. If $\xi_{it}(z) - \xi_{it}(z^\prime) > 0$, there exists a positive spill-over effect. 

Without loss of generality, potential spill-over effects could also arise in different state transitions such as $(1,1) \rightarrow (2,1) \rightarrow (2,2)$ and, hence, the above formulation can be calculated for them analogously.

\subsection{Model Performance}

\textbf{Model fit:} We evaluate the fit of our \CHMM using standard performance metrics for Bayesian modeling \citep{gelman2013bayesian}, that is, the expected log pointwise predictive density (elpd). The elpd can be efficiently computed using the help of Bayesian leave-one-out cross-validation and Pareto smoothed importance sampling \citep{VehtariAki2017PBme}. In our case, we measure model performance for the hold-out sample at patient level (rather than at observation level). For this, we used the \texttt{loo} package as part the statistical software \texttt{R}. Furthermore, we report the widely-applicable information criterion~(WAIC).

\textbf{Model variants:} We compare our proposed \CHMM against alternative model variants. The model variants: (1)~A simplified variant where disease~\A has only a single latent state, \ie, $\big\vert \mathcal{S}^{(A)} \big\vert = 1$. (2)~A simplified variant where disease~\B has only a single latent state, \ie, $\big\vert \mathcal{S}^{(B)} \big\vert = 1$. These represent simplified disease dynamics with only reduced set of states. Note that both model variants have access to identical data as our \CHMM. As we shall see later, the proposed \CHMM with $\big\vert \mathcal{S}^{(A)} \big\vert = \big\vert \mathcal{S}^{(B)} \big\vert = 2$ is superior, thus implying the necessity of accommodating different disease states. 

\subsection{Estimation Procedure}

\textbf{Sampling:} In the \CHMM, all model parameters are obtained via a fully Bayesian estimation \citep{gelman2013bayesian}, specifically through the use of Markov chain Monte Carlo (MCMC) sampling. For this, we used the probabilistic language \texttt{Stan} \citep{CarpenterStan}. Based on it, we obtained posterior estimates from the Hamiltonian Markov chain algorithm together with the No-U-Turn sampler  \citep{HomanNoUTurn}. We ran four chains with 1500 iterations each (1500 additional iterations were discarded as part of a warm-up) to obtain posterior estimates. For this, we first derived the likelihood $\mathcal{L}$ of the \CHMM via
\begin{equation}
\begin{split}
\mathcal{L} = & \prod_{i=1}^N P(y_{it}^{(A)}, \dots,  y_{iT}^{(A)},  y_{it}^{(B)}, \dots,  y_{iT}^{(B)}) = \\
& \prod_{i=1}^N \sum_{s_1=1}^{\vert \mathcal{S}^{(G)} \vert} \sum_{s_2=1}^{\vert \mathcal{S}^{(G)} \vert} \dots \sum_{s_T=1}^{\vert \mathcal{S}^{(G)} \vert} 
\left[ \pi_{s_1} \times \prod_{t=2}^T \gamma^{(s_{t-1} \rightarrow s_t)}_{i,t-1} \times \right. \\ 
& \left. \prod_{t=1}^T
\mathcal{N}(y_{it}^{(A)};  \mu_{\phi^{(A)}(s_t)}^{(A)},  \sigma^{2}_{(A)})
\times \mathcal{N}(y_{it}^{(B)}; \mu_{\phi^{(B)}(s_t)}^{(B)},  \sigma^{2}_{(B)}) \right],
\end{split}
\end{equation}
where $\phi^{(A)}(s_t)$ and $\phi^{(B)}(s_t)$ map the global state to the corresponding individual states of each disease (see Tbl.~\ref{tbl:state_mapping}). We further applied the forward algorithm \citep[][pp.\,36--39]{zucchini2017hidden} in order to obtain an efficient calculation scheme. Moreover, the state-dependent means were ordered, \ie, $\mu_1^{(A)} < \mu_2^{(A)}$, \ldots, $\mu_1^{(B)} < \mu_2^{(B)}$, to avoid label-switching and to promote identifiability \citep[see][]{JasraLabelSwitching}. The runtime on typical hardware amounted to around 12~hours.

\textbf{Priors:} We used weakly informative priors for the emission component and the initial state probabilities:
\begin{align}
\mu_s^{(A)}, \mu_s^{(B)} & \sim \mathcal{N}(0, 10^2), \\
\sigma_{(A)}, \sigma_{(B)} & \sim \mathcal{N}(0, 1), \\
\pi_s & \sim \text{Dirichlet}(1,1,1,1).
\end{align}
Slightly narrow priors were chosen for the intercepts $\alpha_{jk}$ with the aim to stabilize the model fit. Instead of specifying priors for the coefficients $\beta_{jk}$, they were placed on the transformed parameters $\tilde{\beta}_{jk}$, which resulted from a thinned QR decomposition of the centered covariates $x_{it}$. This corresponds to
\begin{align}
\alpha_{jk} & \sim \mathcal{N}(0, 2.5^2), \\
\tilde{\beta}_{jk} & \sim \mathcal{N}(0, 1).
\end{align}

\textbf{Diagnostics:} We follow common guidelines in Bayesian modeling \citep{gelman2013bayesian} and inspect the convergence of the Markov chains. We further conduct posterior predictive checks of whether we can simulate reasonable new observations based on the estimated model parameters \citep{GabryVisualization}. Both are reported as part of the model diagnostics in Sec.~\ref{sec:model_diagnostics}. In addition to that, we checked if our \CHMM was able to recover the specified parameters from entirely simulated data. If not stated otherwise, we later report the posterior mean, as well as the 50\,\% and 90\,\% credible intervals (CrI).

\section{Empirical Setup}
\label{sec:empirical_setup}

\subsection{Data}

We evaluate our \CHMM based on two co-occurring diseases: (\A) diabetes mellitus type 2 and (\B) chronic liver disease. Both often co-occur as comorbidities \citep{Harris115, HICKMAN2007829,Tolman734}. Diabetes is one of the most common chronic disease, impairing the lives of millions of people worldwide. Diabetes further causes substantial cost during the provision of care \citep{world2016global}. On the other hand, chronic liver disease is a medical condition often co-occurring with diabetes  \citep{Harris115, HICKMAN2007829, Tolman734}. In our analysis, we use blood glucose level as lab measurements for diabetes and the level of alanine aminotransferase, an enzyme predominantly found in the liver \citep{Fraser741, Vozarova1889}, as a lab measurement for chronic liver disease.

We later compute spill-over effects following the treatment of increased blood glucose level with Metformin. Metformin is the first line treatment for diabetes type 2 and lowers overall blood glucose levels. In the literature, Metformin is also considered to have a beneficial impact for patients with liver disease \citep{BaileyMetformin, ElisabettaBugianesi2005ARCT, Harris115}. Hence, this raises the question to what extent spill-over effect influence the disease state for chronic liver disease. 

In our analysis, we utilize a longitudinal dataset of patients with prediabetes. The dataset consists of annual electronic health records \citep{AdverseDrugEHR, TASTE_EHR} that also include recordings on lab measurements, treatments, and risk factors (see next section). Overall, the dataset includes 675 patients. Each patient has a time series with recordings between 4 and 10 consecutive years. This amounts to 3253 observations.

\subsection{Model Variables}
\label{sec:model_variables}

Our \CHMM is estimated using the following variables (Tbl.~\ref{tbl:model_variables}).

The observations $Y_{it}^{(A)}$ and $Y_{it}^{(B)}$ correspond to the glucose level and the alanine aminotransferease level as measurements of diabetes and chronic liver disease, respectively. We log-transform both variables due to their right skewed distributions.

The covariates $x_{it}$ characterize between-patient heterogeneity. These represent risk factors and data on treatments. Here we use the body mass index (BMI), sex, age (at time step $t = 1$), and time-since-prediabetes. We explicitly decompose temporal information into both age and time-since-prediabetes as this allows us to isolate age as a risk factor and potential trends.  Data on treatments is given by medication with Metformin. This is a binary variable of whether a Metformin treatment (\ie, Glucomin or Glucophage) was prescribed.  

\begin{table}[H]
\begin{center}
\caption{Overview of model variables.} \label{tbl:model_variables}
\footnotesize
\resizebox{\columnwidth}{!}{
\begin{tabular}{ll}
\toprule
\textbf{Model component} & \textbf{Variables} \\
\midrule
\textbf{Observations} & \\
\quad Diabetes $Y_{it}^{(A)}$ & Blood glucose level (log) \\
\quad Chronic liver disease $Y_{it}^{(B)}$ & Alanine aminotransferase level (log)\\
\midrule
\textbf{Transition mechanism} & \\
\quad \multirow[t]{3}{*}{Covariates $x_{it}$} & BMI, sex, age, time-since-prediabetes, \\
& Metformin treatment (centered), \\
& Metformin treatment (lag=1; centered) \\
\bottomrule
\end{tabular}}
\end{center}
\end{table}

\noindent
The following procedure was used to infer the so-called \emph{within} effect of Metformin. Metformin is generally expected to reduce or stabilize the glucose levels within patients \citep{BaileyMetformin}. However, patients who take Metformin might have higher glucose levels compared to other ones. Thus, the so-called \emph{within} and \emph{between} effects might be conflicting. For that reason, the Metformin variable is centered with the corresponding patient specific mean in order to infer about the \emph{within} effect. This approach is consistent with previous literature \citep{bell_jones_2015, CurranDisaggregation, EndersCraigK.2007CPVi, HamakerCenter, HoffmanPersonsasContexts,WangLijuanPeggy2015ODBa}. Besides the centered Metformin variable, a lagged version of it was also included. These two Metformin variables will be used to estimate potential treatment and spill-over effects over time.

\subsection{Summary Statistics}

Summary statistics of the model variables are given in Tbl.~\ref{tbl:summary_stats}. Blood glucose levels tend to be elevated as compared to recommendations from the World Heath Organization as expected in this population of individuals with \mbox{(pre-)}diabetes. Similarly, alanine aminotransferase values up to 143.40 u/l, suggesting the presence of liver disease in some of the patients \citep[\eg,][]{Harris115}. Overall, patients are characterized by rather high BMI implying a frequent presence of obesity. This is expected as a high BMI is a known risk factor for diabetes. Only a subset of patients received a Metformin treatment. 

We observe a significant correlation between both measurements, namely blood glucose and alanine aminotransferase levels. The Pearson's correlation coefficient amounts to 0.11 ($p < 0.001$). Hence, a higher blood glucose level coincides with a higher alanine aminotransferase level. This points towards potential interactions among both diseases, yet it does not allow for statistical inferences regarding the underlying longitudinal dynamics.

\section{Numerical Results}
\label{sec:results}

\subsection{Model Performance}

The results of the performance comparison are reported in Tbl.~\ref{tbl:model_performance}. The models are evaluated based on the expected pointwise predictive density (elpd). The elpd assesses predictive accuracy in a Bayesian framework and, therefore, the overall fit of a model \citep{gelman2013bayesian, VehtariAki2017PBme}. A higher elpd indicates a better fit. Among the models which converged, the highest elpd (2,148.78) is registered for the proposed \CHMM. It thus achieves the best overall model fit. 

We use baseline models that are suitable for jointly modeling multiple time series. Specifically, we draw upon the baselines from \cite{PohleJennifer2020Apoc}. Alternative models are typically designed for single time series or lack a mechanism for disease interactions (see Sec.~\ref{sec:related_work}). Hence, their choice for the purpose of our paper is precluded.

\emph{Which number of latent states should be selected?} We compare our \CHMM against baselines with a simplified latent state space, that is, a \CHMM where the latent states for either disease \A or \B are removed. Here simplified \A refers to variant with a single latent state for diabetes and, analogously, simplified~\B to a variant with a single latent state for chronic liver disease. As a result, the diseases are assumed to have only a single state (yet which would oppose the Corbin-Straus trajectory framework \citep{Strauss:1988aa,Corbin:1991aa}). Both the simplified~\A (elpd: 1,620.96) and a simplified~\B (elpd: 1,450.83) are outperformed by the proposed \CHMM by a considerable margin. Taken together, this confirms the importance of including latent states and thus accounting for different trajectory phases (\ie, disease states). In subsequent sections, we thus report estimation results from the proposed \CHMM with $2\times2$ states as this model is favored in the model comparison.

\emph{How should between-patient heterogeneity be modeled?} To answer this, we run additional comparisons. First, we draw upon a na{\"i}ve coupled HMM \cite{BrandCoupledHMM}. This model does not use any covariates for modeling between-patient heterogeneity (\ie, neither in the emission component nor in the transition mechanism). Evidently, the model is inferior. Second, we use the emission coupled HMM from Pohle et al.~\cite{PohleJennifer2020Apoc}. This model includes covariates in the emission (whereas our model includes covariates in the transition). However, the chains during MCMC sampling did not converge and, because of this, interpretability for model coefficients is precluded. Moreover, the estimate of the elpd may be unreliable. Third, as a remedy, we modified the emission coupled HMM from Pohle et al.~\cite{PohleJennifer2020Apoc}. Here we additionally introduce an ordering of the estimated means besides ordered intercepts in order to prevent label-switching across latent states, set initial values for intercepts based on a $k$-means procedure \citep{DamianoLuis2018ATOH}, and follow common practice in Bayesian modeling whereby the covariates are subject to a QR decomposition. The latter should remove posterior correlations among covariates and thus facilitate convergence of the MCMC chains \citep{gelman2013bayesian}. To confirm the accuracy of our implementation, we tested the modified emission coupled HMM based on simulated data, where MCMC sampling converged and where the model coefficients were recovered successfully. The model converged successfully for $1\times2$ states (but not for the other state combinations). Overall, the model has a poorer fit compared to our proposed model. Altogether, this implies that that covariates should be included in the transition mechanism (and not in the emission component).

\emph{What is the utility of modeling disease interactions?} By definition, coupled HMMs with only a single latent state cannot account for disease interactions. Hence, we now assess the corresponding benefit. For the na{\"i}ve coupled HMM, we find that model with $2\times 2$ states performs best, implying that the capturing disease interactions is beneficial. (We cannot comment on the emission coupled HMM due to divergence in the chains.) For the proposed \CHMM, we also find that the model with $2\times 2$ states is best. Hence, it must be assumed that both diabetes and chronic liver disease are subject to longitudinal disease interactions. This is later also tested statistically by interpreting the transition probability.

\begin{table*}[ht]
\caption{Model performance comparison.}
\label{tbl:model_performance}
\small
\centering
\begin{tabular}{l cc SSS}
  \hline
 & \textbf{Disease} & \textbf{\#Latent} & \multicolumn{1}{c}{\textbf{elpd}} & \multicolumn{1}{c}{\textbf{SE(elpd)}} & \multicolumn{1}{c}{\textbf{WAIC}} \\ 
 & \textbf{interactions} & \textbf{states} \\
  \hline
  {Na{\"i}ve coupled HMM \cite{BrandCoupledHMM}} & \xmark & (1 $\times$ 2) & 1622.54 & 150.1 & -3245.17 \\
  {Na{\"i}ve coupled HMM \cite{BrandCoupledHMM}} & \xmark & (2 $\times$ 1) & 1430.12 & 140.83 & -2860.39 \\
  {Na{\"i}ve coupled HMM \cite{BrandCoupledHMM}} & \cmark & (2 $\times$ 2) & 2133.57 & 148.64 & -4267.34 \\ 
  {Emission coupled HMM (unmodified) \citep{PohleJennifer2020Apoc}} & \xmark & (1 $\times$ 2) & {(1775.55)$^\dagger$} & {(152.19)$^\dagger$} & {($-$3491.69)$^\dagger$} \\
  {Emission coupled HMM (unmodified) \citep{PohleJennifer2020Apoc}} & \xmark & (2 $\times$ 1) & {(1462.99)$^\dagger$} & {(148.61)$^\dagger$} & {($-$2483.17)$^\dagger$} \\
  {Emission coupled HMM (unmodified) \citep{PohleJennifer2020Apoc}} & \cmark & (2 $\times$ 2) & {(2280.02)$^\dagger$} & {(153.16)$^\dagger$} & {($-$4509.16)$^\dagger$} \\ 
   {Emission coupled HMM (modified) \citep{PohleJennifer2020Apoc}} & \xmark & (1 $\times$ 2) & 1932.06 & 148.19 & -3864.83 \\ 
  {Emission coupled HMM (modified) \citep{PohleJennifer2020Apoc}} & \xmark & (2 $\times$ 1) & {(1383.15)$^\dagger$} & {(142.6)$^\dagger$} & {($-$2528.34)$^\dagger$} \\
  {Emission coupled HMM (modified) \citep{PohleJennifer2020Apoc}} & \cmark & (2 $\times$ 2) & {(1424.63)$^\dagger$} & {(150.14)$^\dagger$} & {($-$2563.72)$^\dagger$} \\ 
  {\CHMM (simplified \A )} & \xmark & (1 $\times$ 2) & 1620.96 & 149.61 & -3241.99 \\
  {\CHMM (simplified \B)} & \xmark & (2 $\times$ 1) & 1450.83 & 139.70 & -2902.97 \\ 
  \textbf{\CHMM (proposed)} & \cmark & (2 $\times$ 2) & \bfseries 2148.78 & 147.29 & \bfseries -4298.73 \\ 
 \hline
 \multicolumn{4}{l}{\scriptsize $^\dagger$ MCMC sampling did not converge} \\
\multicolumn{4}{l}{\scriptsize elpd: expected log pointwise predictive density (higher values are better); SE: standard error} \\ \multicolumn{4}{l}{\scriptsize WAIC: widely-applicable information criterion (lower values are better).} \end{tabular}
\end{table*}

\subsection{Estimation Results}

\textbf{Emission component:} The estimated parameters from the emission component are reported in Tbl.~\ref{tbl:obsv_distr}. For diabetes, the \STABLE state is associated with an average blood glucose level of only $\exp(4.55) = 94.63$ mg/dl. As a expected, the \ACUTE state has a higher value, with a mean of $\exp(4.70) = 109.95$ mg/dl. For both, the credible intervals do not overlap and are thus different at a statistically significant level.

For chronic liver disease, the mean of the \STABLE state amounts to $\exp(2.86)= 17.46$ u/l. This is lower than the mean of the \ACUTE state (30.88 u/l). The credible intervals of do not overlap and, hence, the posterior means of the \STABLE and \ACUTE states are different at the statistically significant level.

\begin{table}[ht]
\begin{center}
\caption{Posterior estimates of emission component.} \label{tbl:obsv_distr}
\small
\begin{tabular}{lcccc}
  \toprule
 & \textbf{Latent state} & \textbf{Mean} & \multicolumn{2}{c}{\textbf{Credible interval}} \\
 \cmidrule(lr){4-5}
 & & & \textbf{5\,\%} & \textbf{95\,\%} \\ 
  \midrule
\multicolumn{5}{l}{\textbf{Diabetes} (blood glucose level; in log)} \\
\quad $\mu_1^{(A)}$ & \STABLE & 4.55 & 4.54 & 4.55 \\ 
\quad $\mu_2^{(A)}$ & \ACUTE & 4.70 & 4.69 & 4.70 \\ 
\quad $\sigma_{(A)}$ & ---- & 0.09 & 0.09 & 0.09 \\ 
\midrule
\multicolumn{5}{l}{\textbf{Chronic liver disease} (alanine aminotransferase level; in log)} \\
\quad $\mu_1^{(B)}$ & \STABLE & 2.86 & 2.85 & 2.88 \\ 
\quad $\mu_2^{(B)}$ & \ACUTE & 3.43 & 3.41 & 3.45 \\ 
\quad $\sigma_{(B)}$ & --- & 0.30 & 0.29 & 0.31 \\ 
   \bottomrule
\end{tabular}
\end{center}
\end{table}

\textbf{Transition mechanism:} Estimates of the parameters regarding the state transition probabilities as follows. Due to high number of coefficients, we refrain from presenting them all and rather focus on how Metformin treatment influences the transition probability. This thus quantifies the direct treatment effect of Metformin on the disease state. Informed by clinical research, we expect the following. (1)~Recall that Metformin is a diabetes treatment. Hence, it should only have an effect on diabetes and not on chronic liver disease. This is confirmed (all corresponding CrI do not include zero). (2)~Metformin has a short-term effectiveness and, hence, we expect to see treatment effects in the non-lagged variable (rather than in the lagged). Again, we yield positive evidence. (3)~We expect a positive treatment effect which would be seen in the transition (\ACUTE,\ACUTE) $\rightarrow$ (\STABLE,\ACUTE). Here a large portion of the probability mass is above zero, indicating a positive treatment effect of Metformin. Here the posterior mean corresponds to an increase of the log-odds of 3.29, resulting in a $\exp(3.29)=26.84$ odds ratio for a one unit increase in the centered Metformin variable and, hence, making the aforementioned transition more likely than staying in the state (\ACUTE,\ACUTE).

\textbf{Coupling:} We now investigate the coupling among the two comorbidities. For that, we report the baseline probability in the transition mechanism (\ie, the intercept $\alpha_{jk}$). It quantifies the transition conditional on the latent states of both disease \A \emph{and} \B. To estimate this, all covariates were set to respective sample mean. This corresponds to $x_{it}=\overline{x}_{it}$ in Eq.~\ref{eqn:tpm_eta}. 

\begin{figure}[h]
\centering
\includegraphics[width=\columnwidth]{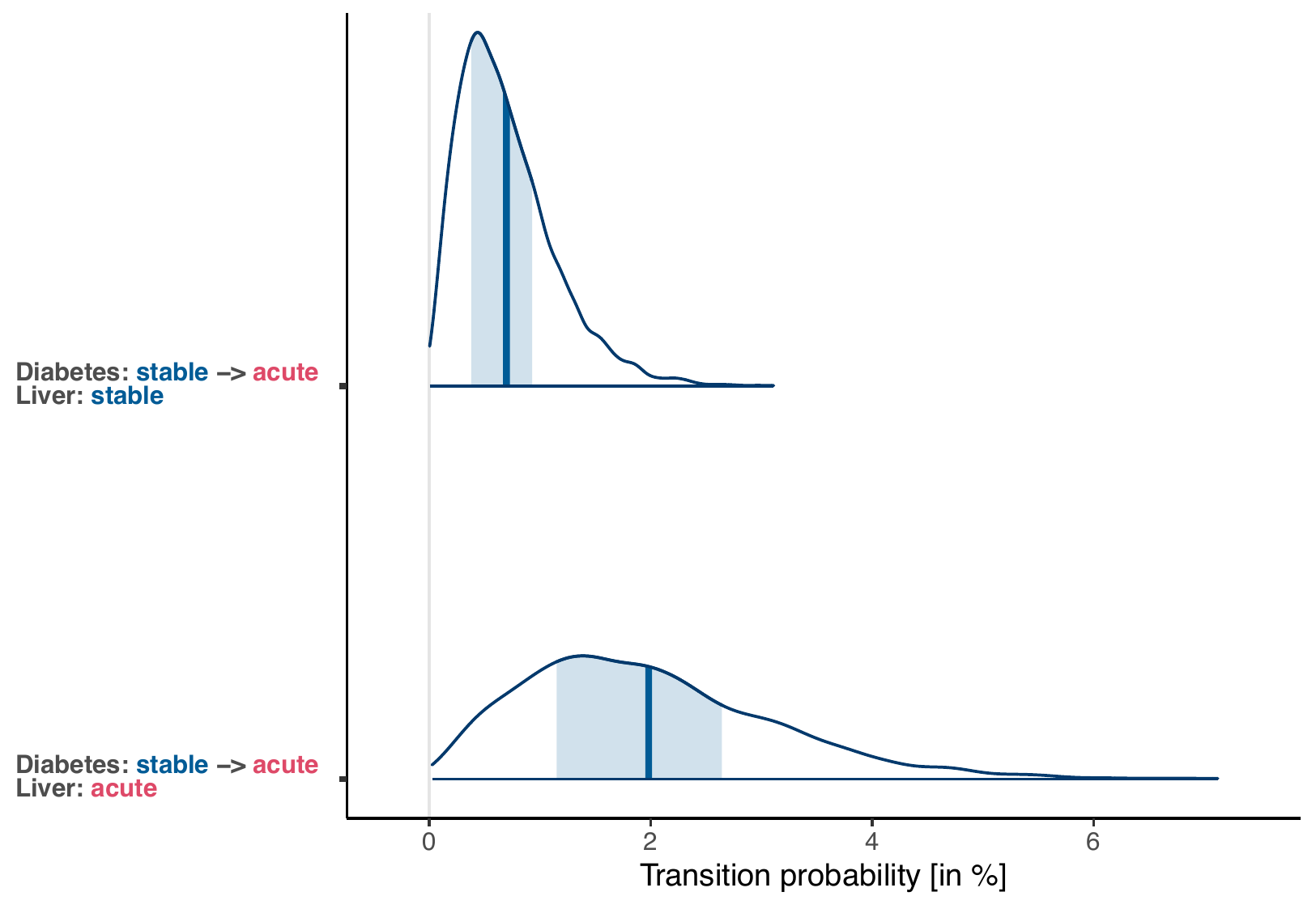}
\begin{flushleft}
\scriptsize\emph{Note:} Here shown are the transitions probabilities that diabetes moves from a \STABLE to an \ACUTE latent state, \ie, $(1,\square) \to (2,\square)$. Hence, this compares the transition probability when exiting a \STABLE chronic liver disease (top) and an \ACUTE chronic liver disease (bottom). Estimations show the posterior means (thick line), as well as the 50\,\% (shaded area) credible interval. All covariates are set to the respective sample mean during estimation. 
\end{flushleft}
\caption{Estimation results for coupling conditional on diabetes.}
\label{fig:gluc_transitions}
\end{figure}

Fig.~\ref{fig:gluc_transitions} shows the coupling for diabetes. The plot gives the probability of disease~\A (diabetes) to transition from a \STABLE to an \ACUTE latent state \emph{conditional} on the latent state of both diseases. If both disease are independent, then the credible intervals for the conditional transition probability should include zero. Larger conditional transition probability values indicate a stronger coupling. From Fig.~\ref{fig:gluc_transitions}, we find that an \ACUTE state for chronic liver disease is associated with an increased probability of moving from \STABLE to \ACUTE in diabetes; see $(1,2) \to (2,2)$. Put simply, if chronic liver disease is \ACUTE, then it is likely that diabetes also transitions from \STABLE to \ACUTE. Overall, we find strong evidence that diabetes is coupled with chronic liver disease.  

\begin{figure}[h]
\centering
\includegraphics[width=\columnwidth]{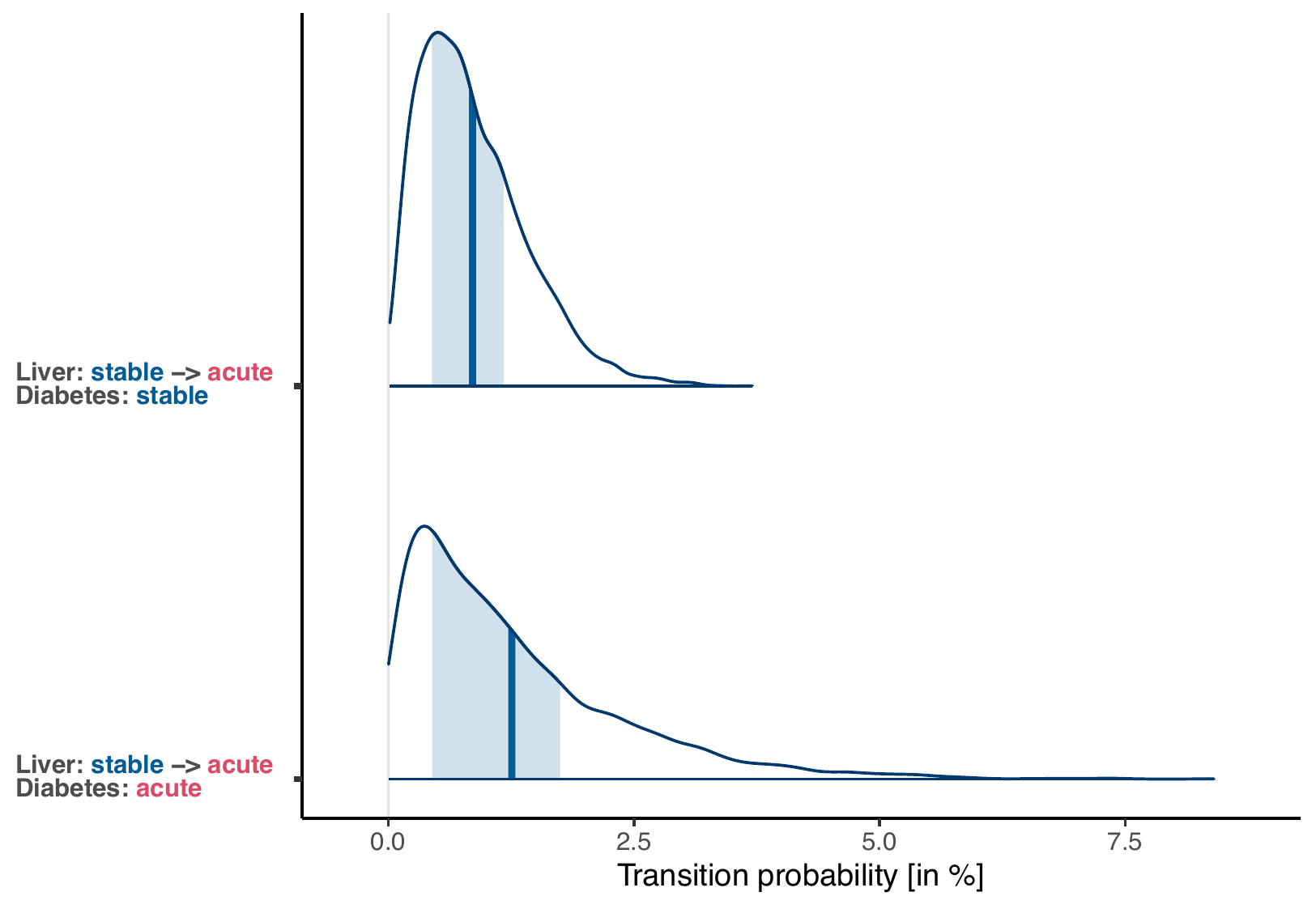}
\begin{flushleft}
\scriptsize\emph{Note:} Here shown are the transitions probabilities that chronic liver disease moves from a \STABLE to an \ACUTE latent state, \ie, $(\square,1) \to (\square,2)$. Hence, this compares the transition probability when exiting a \STABLE diabetes (top) and an \ACUTE diabetes (bottom). Estimations show the posterior means (thick line), as well as the 50\,\% (shaded area) credible interval. All covariates are set to the respective sample mean during estimation. 
\end{flushleft}
\caption{Estimation results for coupling conditional on chronic liver disease.}
\label{fig:altg_transitions}
\end{figure}

The interpretation holds analogously when exchanging \A and \B. Here similar effects can also be detected in Fig.~\ref{fig:altg_transitions}. Put simply, if diabetes is already \ACUTE, then it is somewhat likely that chronic liver disease also becomes \ACUTE. However, the effect is less pronounced and the 50\% and 90\% credible interval overlap to a great extent. Nevertheless, these findings suggest that the two comorbidities indeed interact with each other.

\subsection{Estimated Spill-Over Effect}

Lastly spill-over effects are presented. Specifically, we assess the effect of Metformin on liver disease through the co-evolution of the two comorbidities. For this, we consider the state transitions $(2,2) \rightarrow  (1,2) \rightarrow  (1,1)$. It reflects the transition from \ACUTE to \STABLE in diabetes followed by the transition from \ACUTE to \STABLE in liver disease. In particular, we report the following: (1)~Metformin has been prescribed, \ie, $\xi_{it}(z)$. Here the centered variable is set to 0.5, whereby the remaining covariates are also assumed to be equal to the sample mean. (2)~No Metformin has been prescribed, \ie, $\xi_{it}(z')$. This corresponds to a centered Metformin variable equal to zero and all remaining covariates equal to the respective sample mean. (3)~The spill-over effect, which amounts to the absolute difference between both. (4)~A quotient. It denotes how many times more likely a spill-effect is when comparing Metformin vs. no Metformin. The corresponding quantiles of the transition probabilities of these two scenarios as well as their difference are reported in Tbl.~\ref{table_spill_over}.

Overall, we find the following. If no Metformin is prescribed, there is hardly any change in the state for liver disease through the underlying coupling. Here all estimated quantiles are close to zero. If Metformin is prescribed, we observe a non-zero probability that the state from chronic liver disease also improves. By comparing the different quantiles, we observe that the distribution for $\xi_{it}(z)$, \ie, Metformin prescribed, has a very long tail. For many patients on the left-tail, the spill-over effect is small or absent (\eg, due to non-adherence or simply because the treatment is not effective). Based on our results, it cannot be ruled out that, for the 5\,\% quantile, the Metformin treatment has also negative implications for a co-occurring liver disease. 

However, for the patients on the right-tail, there is a large estimated spill-over effect. For instance, for the 50\,\% quantile, prescribing Metformin (as opposed to non-prescribing) makes a improvement in the liver state 2.898 times more likely. In the 75,\% quantile, the relative quotient increases even to 13.695. As shown here, our model allows to quantify the spill-over effect of diabetes on liver disease due to the use of Metformin. In sum, we find evidence that Metformin stabilizes not only diabetes but, through the coupling, also a co-occurring chronic liver disease.

\begin{table}[ht]
\begin{center}
\caption{Estimated spill-over effects.} \label{table_spill_over}
\resizebox{\columnwidth}{!}{
\begin{tabular}{lrrrrr}
  \toprule
 & \textbf{5\%} & \textbf{25\%} & \textbf{50\%} & \textbf{75\%} & \textbf{95\%} \\ 
  \midrule
Metformin prescribed ($\xi_{it}(z)$) & 0.000\,\% & 0.002\,\% & 0.017\,\% & 0.089\,\%& 0.627\,\% \\ 
No Metformin prescribed ($\xi_{it}(z')$) & 0.000\,\% & 0.002\,\% & 0.006\,\% & 0.016\,\% & 0.044\,\% \\ 
\midrule
Difference & $-$0.019\,\% & $-$0.002\,\% & 0.008\,\% & 0.077\,\% & 0.619\,\% \\ 
Quotient & 0.022 & 0.463 & 2.898 & 13.695 & 91.276 \\ 
   \bottomrule
\end{tabular}}
\caption*{\scriptsize{Note: The quantiles of the probabilities of the transition $(2,2) \rightarrow  (1,2) \rightarrow  (1,1)$ are reported. Metformin prescribed assumes that the centered Metformin variables is equal to 0.5 and the remaining covariates are equal to the sample mean. In contrast, no Metformin prescribed corresponds to a centered Metformin variable equal to 0 and the remaining covariates also set to the sample mean. Additionally, the quantiles of the difference of these transition probabilities are given.}}
\end{center}
\end{table}

\section{Discussion}
\label{sec:discussion}

\textbf{Summary of findings:} (1)~Our results find that our \CHMM outperforms alternative model specifications. In particular, it is superior over variants where no disease interaction is modeled. Here we obtain substantial improvements in the model performance as measured by predictive accuracy for Bayesian modeling (\ie, the expected pointwise predictive density, elpd). (2)~We provide empirical evidence that both diabetes and chronic liver disease are dependent. This is revealed by the fact that transitions of one disease depend on the state of the other disease at a statistically significant level. Thereby, we confirm the importance of coupling in our model formation. (3)~Our model returns empirical evidence characterizing the spill-over effect of a diabetes treatment on the state from chronic liver disease.  

Prior literature suggested mixed findings on the role of Metformin for patients with chronic liver disease \citep{BaileyMetformin, ElisabettaBugianesi2005ARCT, Harris115}, especially as a direct effect would not necessarily be supported by the molecular mechanism of action. Here we contribute empirical evidence whereby Metformin has a direct effect on the state for diabetes, while we find no support for a direct effect on the state for chronic liver disease. Instead, we find an indirect effect through the coupling.

\textbf{Interpretability:} Our model is based on a parsimonious specification in order to warrant for interpretability. Here we build upon a central concept from medicine: diseases form a so-called trajectory which undergoes different phases with \ACUTE and \STABLE disease states \citep{Strauss:1988aa,Corbin:1991aa}. These disease states are directly relevant for clinical decision-making as different treatment plans should be applied depending on whether a patient is in an \ACUTE vs. a \STABLE state. Specifically it suggests that clinical decision-making should tailor care to the underlying disease state (and not measurements or symptoms as these are only noisy observations of the state) \citep{larsen2017lubkin}. Owing to this, prior research has modeled health trajectories through the use of latent states \citep[\eg,][]{SukkarDiseaseprogression, oezyurt2021attdmm, ShirleyALCOHOLISM, MartinoHMMdisease, BartolomeoNicola2011Polc, ScottHMMLongitudinal}. In line with this, we also introduce latent disease states and use them for coupling the states of different diseases, and, on top of that, we define our spill-over effect with regard to latent states (as opposed to measurements or symptoms). 

\textbf{Limitations:} As with other research in medicine, ours is not free of limitations and we thus make suggestions for future research. First, our data comprises a unique, large-scale collection of electronic health records from a Western health provider. Future research could replicate our approach with other cohorts. Second, we used an extensive set of risk factors informed by prior literature. Nevertheless, other variables might also represent risk factors and, hence, future research could test how the model performance changes by including them. Third, our analysis is based on specific setting, namely the interaction among diabetes and chronic liver disease under Metformin treatment. Here future research could validate our model based on other comorbidities. If desired, the model can also be extended to comorbidities comprising of three (or more) diseases by expanding the underlying latent state space.  

\textbf{Implications:} Our model offers insights for clinical practice and research. By inferring spill-over effects, one can assess to what extent treatments for disease~\A also change the course (disease state) of disease~\B. Hence, when seeking to stabilize the course of disease~\B, clinical practitioners can thus decide upon a treatment with a direct effect on \B or, by contrast, choose a treatment plan for \A that also has an indirect effect such that disease~\B is stabilized. Here our model offers statistical insights, allowing for evidence-based decisions regarding treatment planning.   

Our model was tested based on diabetes and chronic liver disease. However, our model is designed in a generic manner in order to ensure widespread applicability. The clinical co-author from our work foresees several use cases: Here our model could analyze comorbidities that are poorly understood or where there are mixed findings in the clinical literature. Given the vast number of comorbidities, this gives rich opportunities for follow-up research.

\section{Conclusion}
\label{sec:conclusion}

Comorbidities, defined as the presence of multiple co-occurring diseases, are widespread among patients, and yet comprehensive statistical frameworks for modeling the longitudinal dynamics of comorbidities are rare. Such models would allow for novel insights into the co-evolution of comorbidities over time. Specifically, this would allow to answer the question: \emph{how does treating disease~\A change the state of disease~\B?} To address this, we developed a probabilistic model for longitudinal data analysis: a coupled hidden Markov model with personalized, non-homogeneous transitions (\CHMM). To the best of our knowledge, our \CHMM is the first statistical model that is specifically tailored to capture longitudinal dynamics among comorbidities. Our model further ensures interpretability by offering clinically-relevant inferences. In particular, we presented a spill-over effect that measures the indirect effect of a treatment on disease states through the co-development of comorbidities. We evaluated our model based on a longitudinal dataset from 675 patients. We found that \CHMM outperforms alternative model specifications in terms of model fit.

\appendix

\begin{acks}
Financial supported from the Swiss National Science Foundation (SNSF) through an Eccellenza grant (186932) is acknowledged.
\end{acks}

\bibliographystyle{ACM-Reference-Format}
\bibliography{literature}

\clearpage
\newpage
\appendix
\raggedbottom

\section{Additional Tables}
\label{sec:additional_tables}

\begin{table}[h]
\caption{Transition matrix (with corresponding $\alpha$).} \label{tbl:tpm}
\begin{center}
\begin{tabular}{p{0.5cm} p{0.5cm} p{0.1cm} cccc} 
\toprule
\multicolumn{2}{c}{\textbf{Latent states}} & \multirow{2}{*}{\textit{\underline{to}}}   & 1 & 2 & 3 & 4  \\
\multicolumn{2}{c}{\textit{\underline{from}}} &  & (1,1) & (1,2)  & (2,1) & (2,2)    \\ 
\midrule
1 & (1,1) & & --  & $\alpha_{12}$  & $\alpha_{13}$  & $\alpha_{14}$  \\
2 & (1,2) & & $\alpha_{21}$  & --  &   $\alpha_{23}$ &  $\alpha_{24}$ \\
3 & (2,1) & & $\alpha_{31}$ &  $\alpha_{32}$ & -- & $\alpha_{34}$   \\
4 & (2,2) & & $\alpha_{41}$ & $\alpha_{42}$  &  $\alpha_{43}$ &  -- \\
\bottomrule
\end{tabular}
\end{center}
\end{table}

\begin{table}[h]
\begin{center}
\caption{Summary statistics.} \label{tbl:summary_stats}
\centering
\footnotesize
\begin{tabular}{lr}
\toprule
Model variable & Overall ($N=3253$)\\
 \midrule
\textbf{Blood glucose level} [in mg/dl] & \\
\quad Mean (SD) & 99.59 (11.68)\\
\quad Range & 62.00 -- 211.00\\
\textbf{Alanine aminotransferase level} [in u/l] & \\
\quad Mean (SD) & 24.23 (11.43)\\
\quad Range & 4.05 -- 143.40\\
\midrule
\textbf{BMI} & \\
\quad Mean (SD) & 30.55 (5.86)\\
\quad Range & 15.48 -- 66.67\\
\textbf{Sex} & \\
\quad Male & 1194 (36.7\%)\\
\quad Female & 2059 (63.3\%)\\
\textbf{Age} [in years] & \\
\quad Mean (SD) & 55.34 (8.27)\\
\quad Range & 24.00 -- 70.00\\
\textbf{Metformin treatment} & \\
\quad Not prescribed ($=0$) & 3080 (94.7\%)\\
\quad Prescribed ($=1$) & 173 (5.3\%)\\
\bottomrule
\multicolumn{2}{l}{\tiny Omitted (for brevity): time-since-prediabetes, Metformin (lag=1)} \\
\multicolumn{2}{l}{\tiny SD: standard deviation}
\end{tabular}
\end{center}
\end{table}

\section{Model Diagnostics}
\label{sec:model_diagnostics}

\begin{figure}[H]
\centering
\includegraphics[width=\columnwidth]{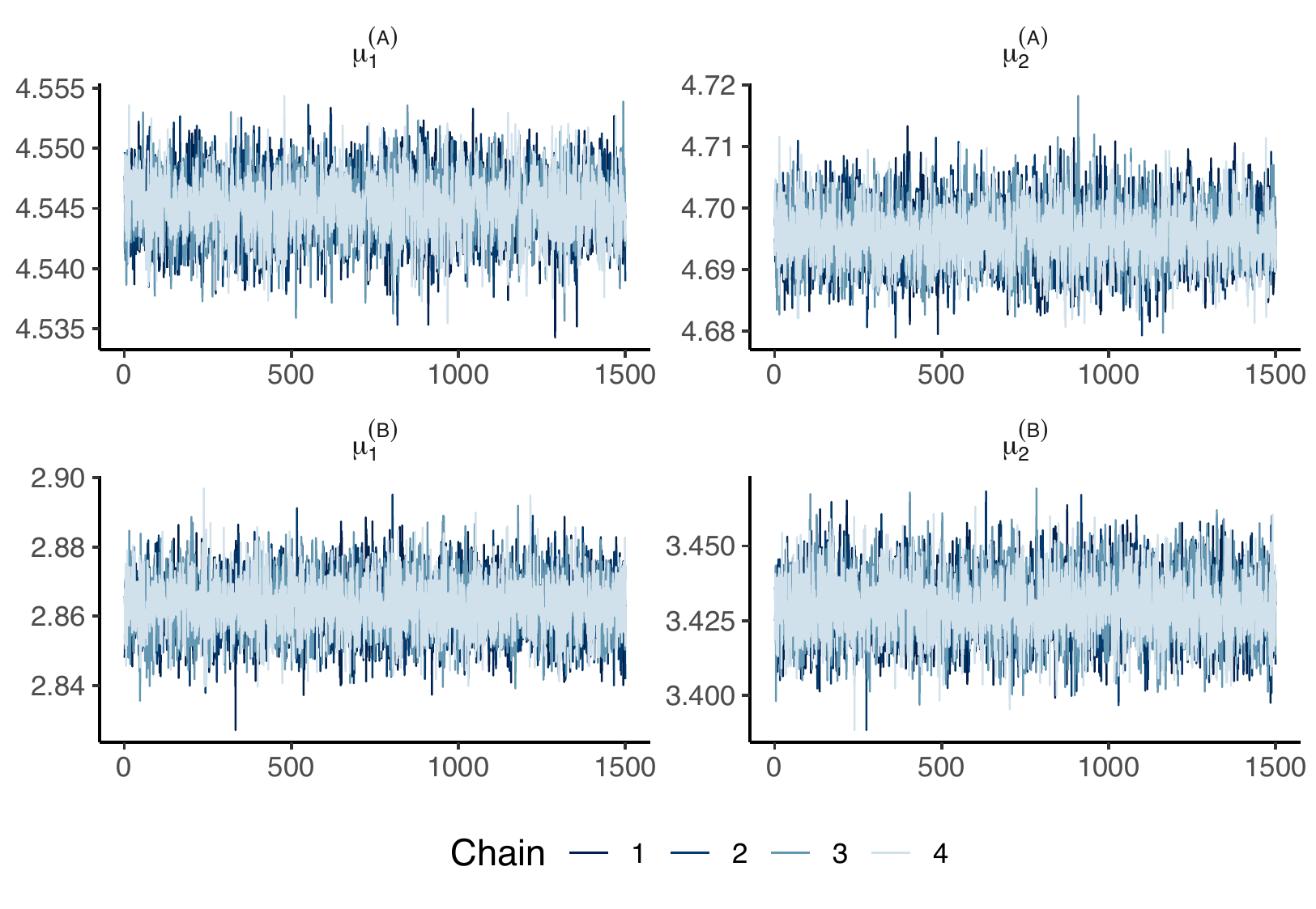}

\vspace{-0.2cm}
\begin{flushleft}
\scriptsize{\emph{Note:} Traceplot for 1500 iterations of each chain (warmup of MCMC algorithm omitted).}
\end{flushleft}
\vspace{-0.2cm}
\caption{Traceplots of emission means $\mu_s^{(A)}$ and $\mu_s^{(B)}$.}
\label{fig:trace_mu}
\end{figure}

We follow common practice in Bayesian modeling \citep{gelman2013bayesian} by performing the following model diagnostics. This is to ensure convergence of the MCMC algorithm and thus precise estimates. First, we inspected the effective sample size $n_\text{eff}$, indicating that the number of MCMC samples is sufficient. Second, we calculated the Gelman-Rubin convergence diagnostic $\hat{R}$ of all model parameters. The $\hat{R}$ is below the critical threshold of 1.1, suggesting convergence of the MCMC chain. Third, we retrieved traceplots (Fig.~\ref{fig:trace_mu}). The traceplots show the MCMC draws of the emission mean across the latent states and difference diseases (\ie, $\mu^{(A)}_s$ and  $\mu^{(B)}_s$, $s \in \{ 1, 2 \}$). The traceplots suggest that the chains have mixed well. Altogether, the previous diagnostics provide ample evidence that the MCMC algorithm has converged.

\begin{figure}[H]
\centering
\subfigure[Observation $Y_{it}^{(A)}$: Blood glucose level (in log)]{\includegraphics[width=\columnwidth]{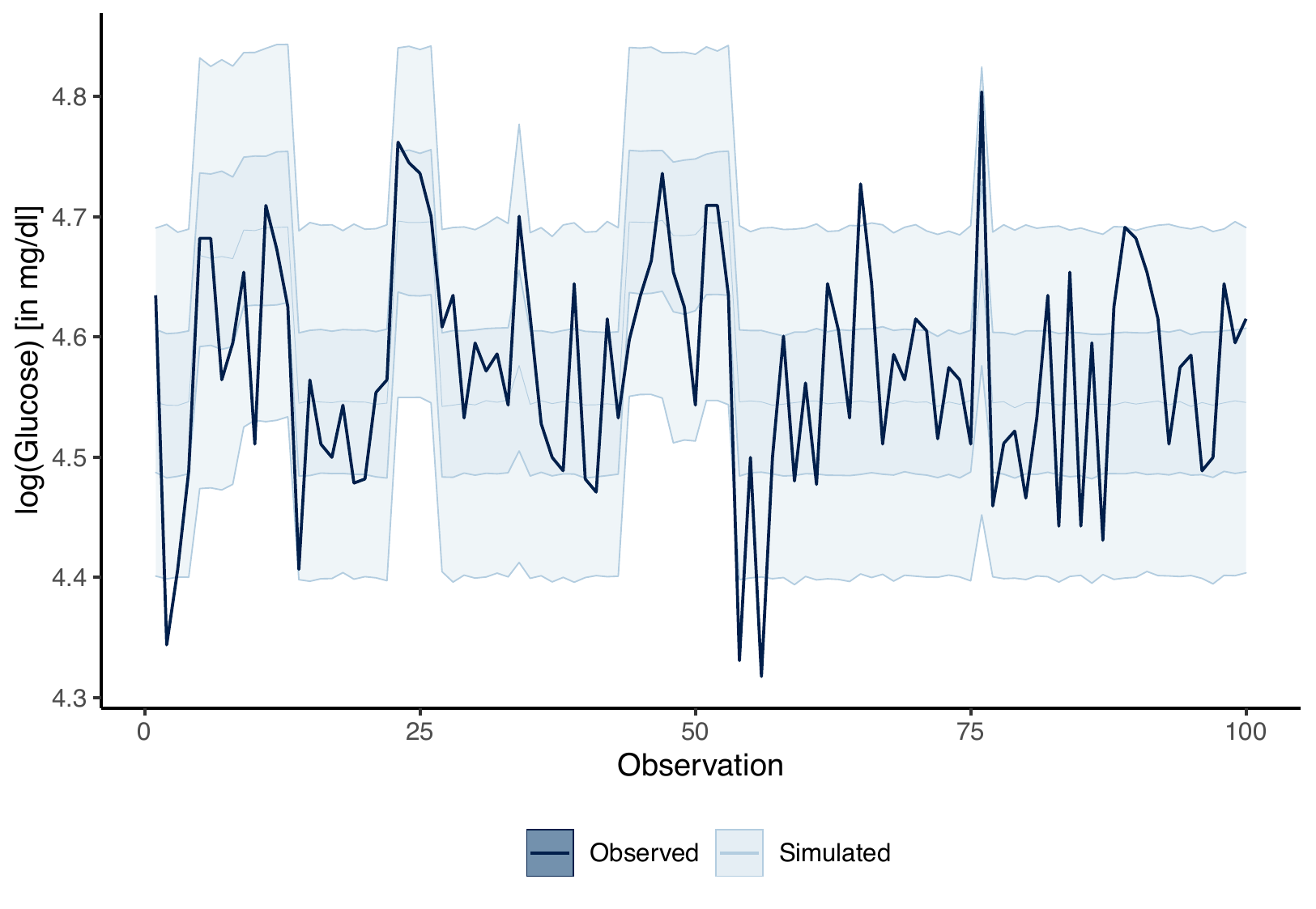}\label{fig:ppc_gluc_2}} \\
\subfigure[Observation $Y_{it}^{(B)}$: Alanine aminotransferase level (in log)]{\includegraphics[width=\columnwidth]{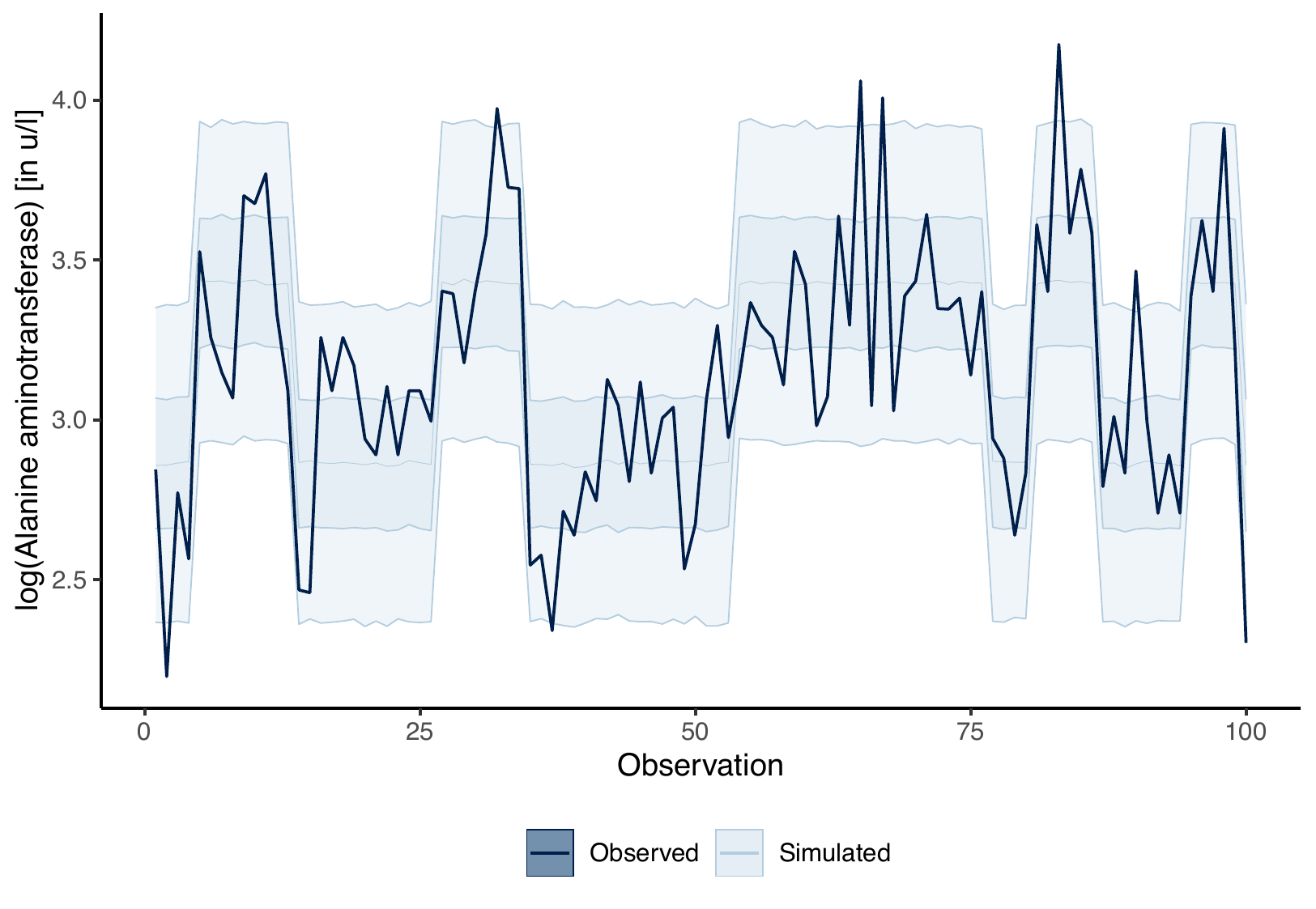}\label{fig:_ppc_altg_2}}

\begin{flushleft}
\scriptsize\emph{Note:} The plots compare the 50\,\% (light color) and 90\,\% (dark color) credible intervals of simulated observations against the actual observations. Reported are blood glucose levels (top) and alanine aminotransferase levels (bottom).
\end{flushleft}

\caption{Posterior predictive checks (indicating the accuracy of the model on simulated data).}
\label{fig:ppc}
\end{figure}

We further assessed our model by conducting posterior predictive checks following recommendations on Bayesian modeling \citep{gelman2013bayesian, GabryVisualization}. For this, new observations were simulated based on the posterior draws of the model parameters as well as the most probable latent states which were estimated using the Viterbi algorithm. The results are in Fig.~\ref{fig:ppc}. Based on it, we compare the credible intervals of the generated observations against the actual observations. The results suggest that our model fits the data well. We further checked whether \CHMM is able to recover the specified parameters from simulated data. Here we obtain confirmatory results, suggesting that it can successfully capture the underlying dynamics. 

\end{document}